\documentclass[conference]{IEEEtran}
\IEEEoverridecommandlockouts
\usepackage{cite}
\usepackage{amsmath,amssymb,amsfonts}
\usepackage{algorithmic}
\usepackage{graphicx}
\usepackage{textcomp}
\usepackage{xcolor}
\usepackage[short]{optidef}
\usepackage{multicol}
\usepackage{multirow}
\usepackage{bm}

\def\BibTeX{{\rm B\kern-.05em{\sc i\kern-.025em b}\kern-.08em
    T\kern-.1667em\lower.7ex\hbox{E}\kern-.125emX}}
\begin{document}

\title{Data-driven HVAC Control Using Symbolic Regression: Design and Implementation}

\author{\IEEEauthorblockN{Yuki Ozawa\IEEEauthorrefmark{1}, Dafang Zhao\IEEEauthorrefmark{1}, Daichi Watari\IEEEauthorrefmark{1}, Ittetsu Taniguchi\IEEEauthorrefmark{1},\\ Toshihiro Suzuki\IEEEauthorrefmark{2}, Yoshiyuki Shimoda\IEEEauthorrefmark{3}, Takao Onoye\IEEEauthorrefmark{1}}
  \IEEEauthorblockA{\IEEEauthorrefmark{1}Graduate school of Information Science and Technology, Osaka University, Japan\\\IEEEauthorrefmark{2}Campus Sustainability Office, Osaka University, Japan\\
  \IEEEauthorrefmark{3}Graduate School of Engineering, Osaka University, Japan}
E-mail: \IEEEauthorrefmark{1}\IEEEauthorrefmark{2}\IEEEauthorrefmark{3}\{ozawa.yuki@ist, zhao.dafang@ist, watari.daichi@ist, i-tanigu@ist,\\ suzuki-to@mail,  shimoda@see.eng, onoye@ist\}.osaka-u.ac.jp}

\maketitle

\begin{abstract}
The large amount of data collected in buildings makes energy management smarter and more energy efficient.
This study proposes a design and implementation methodology of data-driven heating, ventilation, and air conditioning (HVAC) control.
Building thermodynamics is modeled using a symbolic regression model (SRM) built from the collected data.
% This model was implemented and verified in real-world residential space. 
%The temperature response obtained from the proposed model was in good agreement with the observations, with a maximum deviation of less than **$^{\circ}C$. 
Additionally, an HVAC system model is also developed with a data-driven approach. 
A model predictive control (MPC) based HVAC scheduling is formulated with the developed models to minimize energy consumption and peak power demand and maximize thermal comfort.
%On the basis of the above models, the online HVAC control framework is further developed. 
The performance of the proposed framework is demonstrated in the workspace in the actual campus building. 
The HVAC system using the proposed framework reduces the peak power by 16.1\% compared to the widely used thermostat controller.
\end{abstract}


\section{Introduction}
Energy management systems for buildings play an important role in the transition to a low carbon and energy-saving society that has been underway around the world in recent years.
Here, it is known that heating, ventilation, and air conditioning (HVAC) systems are particularly important in building energy management.
HVAC accounts for around 40\% of the total energy consumption of a building, and by operating HVAC without waste based on information such as the number of occupants and the outside temperature, energy savings can be expected for the entire building while maintaining comfort.
In addition, HVAC operation tends to be concentrated in the early mornings in winter and summer.
This concentrated power demand places a heavy burden on the power grid and caused peak power demand.
\textcolor{black}{Much research has been conducted on model predictive control (MPC), one of the promising schemes for HVAC system management.
The MPC approach iteratively computes optimal control input based on system prediction models and forecasting information at every time step.
Ostadijafari et al.~\cite{Ostadijafari2019-at} investigated a non-linear economic MPC for a smart building that includes HVAC systems, PV generation, flexible appliances and a battery system.
Their work's goal is to minimize electricity costs under time-dependent price signals while tracking comfort temperature ranges based on occupancy information.
Cui et al.~\cite{Cui2017-gw} focused on an MIP-based receding horizon approach to optimize battery and HVAC operations. However, they only minimized the system costs consisting of electricity costs and battery degradation costs and did not consider thermal comfort.
% Perez et al.~\cite{Perez2016-hl} proposed an integrated MPC scheme to minimize peak demand by scheduling HVAC and time-shiftable appliances.
% Their main concern was to shave the power peak and did not take comfortable indoor temperature tracking into account.
Chanthawit et al.~\cite{anuntasethakul2021design} designed a model predictive controller for HVAC to minimize operating cost in accounts of peak-shaving and thermal comfort.
Against this background, HVAC management systems are expected to suppress peak power demand and maintain a reasonable level of thermal comfort.
Moreover, most of these works are only theoretical studies or simulations, but have not been tested in the real world.}

Predicting thermal changes in buildings is important for HVAC management systems.
Room temperature prediction methods can be broadly classified into physics-based white-box approaches, data-driven black-box approaches, and gray-box approaches that are hybrids of white- and black-box approaches \cite{BOURDEAU2019101533}.
White-box approaches, such as EnergyPlus \cite{Crawley2001-zk}, require detailed information on building and HVAC systems.
% modeling of the building materials and HVAC layout of the target building.
The gray-box approach, on the other hand, uses historical data from the system to determine the parameters of the physical model, such as the thermal equivalent circuit model~\cite{zhao, Ferracuti2017}.
Both of these approaches require a deep understanding of building thermal modeling.
On the other hand, the black-box only requires historical or real-time data to estimate the thermal dynamics behavior of the building, and there is no need to know the complex physical processes that govern the thermal dynamics of the building~\cite{lundberg2017unified}.
% approach is built by fitting the model to the historical data.
% Although this approach does not require domain knowledge when building the model, it is known to be difficult to interpret the model 
In recent years, symbolic regression has been used to investigate the thermal response of the buildings~\cite{Leprince2021}.
% Here, research has been conducted to investigate the factors that influence the thermal response of buildings using symbolic regression \cite{Leprince2021}.
Symbolic regression is a regression analysis that seeks the exact, simple, and most suitable function for the given data.
Furthermore, since symbolic regression identifies the functions that describe the model, the constructed model is easy to interpret and can effectively determine the physical model using the gray-box approach.
% Symbolic regression is a type of regression analysis that seeks the exact, simple, and most appropriate function for given data.
% The function sought in symbolic regression does not have a standard formula form, but uses genetic programming to search for the best combination of operators, variables, and coefficients.
% This corresponds to the black box approach described above, which does not require domain knowledge when constructing the model.
% In addition, since symbolic regression determines the function that describes the model, the constructed model is easy to interpret and is effective in determining the physical model in the gray box approach.

Therefore, this study proposes the use of symbolic regression as a method of predicting room temperature in online HVAC control.
The goal of this study is to incorporate room temperature prediction using symbolic regression into an online HVAC optimization framework~\cite{zhao} and implement it in a real environment.
Additionally, we propose a novel HVAC model using piecewise-linear functions in order to describe HVAC behavior in more detail.

The rest of this paper is organized as follows. 
Section~\ref{sec:overview} presents an overview of the proposed air conditioning framework.
Section~\ref{sec:SRM} describes the prediction of room temperature using symbolic regression.
The proposed HVAC model is shown in Section~\ref{sec:HVAC}.
Section~\ref{sec:exp} shows the results of the on-site experiment to evaluate the proposed framework.
%describes the implementation of the framework and the experimental results from a real case study.
Finally, a conclusion is presented in Section~\ref{sec:conclusion}.

\section{Overview of HVAC control framework}\label{sec:overview}
In this section, we give an overview of the proposed framework, which uses the MPC approach for HVAC control~\cite{en11030631}.
In the MPC approach, an optimization problem is solved at each time step using predictive information such as ambient temperature and solar irradiation and mathematical models that predict the future behavior of control targets such as room temperature and HVAC system.
The solution contains an HVAC schedule for a given horizon, such as several hours, but only the first step of the solution is used as a control input.
After a given interval, the control horizon moves in one step and the same procedure is repeated to find the optimal control input.
In this way, the control inputs can be updated in real-time to reflect the latest state of the system.

\begin{figure}[!t]
    \centering
    \includegraphics[width=0.8\linewidth]{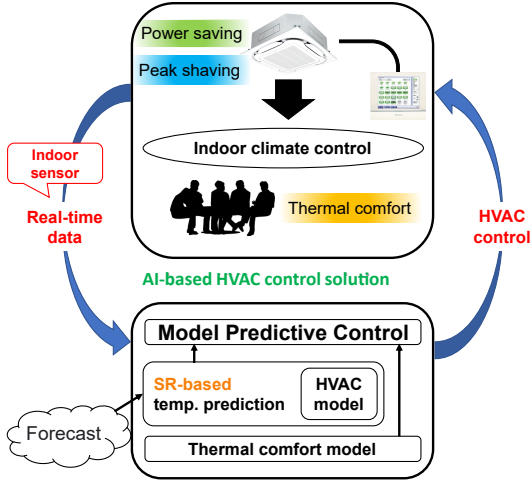}
    \caption{Overview of HVAC control framework.}
    \label{fig:overview}
\end{figure}

Overview of the proposed control framework as illustrated in Fig. \ref{fig:overview}.
The real-time HVAC operation data along with weather forecast data as the input of the optimization problem solver; the output is the HVAC operation schedule which changes the HVAC temperature setpoint to achieve the optimal power consumption.
The proposed framework is designed to minimize the following in the optimal HVAC operational schedule.
\begin{itemize}
    \item Total energy consumption for the control period.
    \item The difference between the most comfortable and the actual room temperature.
    \item HVAC power ramp rate for peak power demand reduction.
\end{itemize}
We briefly describe the optimization problem solved in the framework as follows; please refer to \cite{zhao} for the detailed mathematical formulation:
\begin{mini!}<b>
    {D_t,s_t}{\sum^{T}_{t=0}D_t + \sum^{T}_{t=0}o_t \cdot (T^{in}_t - T^{comf})^2 \label{eq:obj}}{}{}
    \breakObjective{+ \Delta D_{max} + P_e\sum^{T}_{t=0}s_{t} \nonumber}
    \addConstraint{0}{\leq s_t \label{eq:slack}}{\ \ \forall t}
    \addConstraint{\Delta D_t}{ = D_{t+1}-D_{t}\label{eq:deltad}}{\ \ \forall t}
    \addConstraint{\gamma_{t} \cdot \Delta D_{t}}{\leq \Delta D_{max} \label{eq:dmax}}{\ \ \forall t}
    \addConstraint{T^{lb}-s_t}{\leq T^{in}_t\leq T^{ub}+s_t\label{eq:tmpconst}}{\ \ \forall t}
    \addConstraint{T^{in}_{t+1}}{=f^{SRM}(\bm{X}_t)\label{eq:roomtemp}}{\ \ \forall t}
    \addConstraint{Q_{t}}{=f^{HVAC}(\bm{X}_t)\label{eq:hvac}}{\ \ \forall t}
\end{mini!}
where $t$ is the time index ($0\leq t \leq T$), $D_{t}$ is the HVAC power consumption, $s_{t}$ is the slack variable constrained by (\ref{eq:slack}), which are decision variables.
For other variables, $\Delta D_{max}$ is the maximum power ramp rate, $T^{in}_{t}$ is the room temperature, $\Delta D_t$ is the power ramp rate calculated by (\ref{eq:deltad}), and $Q_{t}$ is the thermal capacity provided by HVAC systems.
For parameters and input, $o_t$ is the occupancy information, when occupied, $o_{t}=1$: otherwise 0, $P_e$ is the penalty coefficient for peak power, $\gamma_t$ is the weight for the ramp rate: when peak hours, it is set to $>1$; otherwise 1.
The set point $T^{comf}_{t}$ means the comfortable temperature and $T^{ub}$ and $T^{lb}$ denote the upper and lower bounds of the comfortable temperature range.

In the objective function (\ref{eq:obj}), the first term means the total energy consumption, the second term is the deviation between the room and the comfortable temperature, the third term reduces the maximum ramp rate found by (\ref{eq:dmax}), and the fourth term minimizes the violation of temperature.
Note that we normalize these objective terms to align the scales that the values can take, although it does not appear in the formulation due to space limitations.
This normalization is useful for dealing with objective terms of different scales in a single objective function, and the details are presented in the literature \cite{zhao}.
On the other hand, the constraint (\ref{eq:tmpconst}) ensures that the room temperature does not exceed the comfortable temperature range.
If temperature violation is unavoidable, the slack variable $s_{t}$ relaxes the upper/lower temperature bounds.

In this paper, we develop data-driven functions to predict the change in room temperature and the thermal gain of HVAC.
Using the explanatory variable $\bm{X}_t$ that includes room temperature, ambient temperature, power consumption, etc., the function $f^{SRM}$ in (\ref{eq:roomtemp}) calculates the next step room temperature, and $f^{HVAC}$ in (\ref{eq:hvac}) estimates the HVAC capacity.
It should be stressed that the main contribution compared to work \cite{zhao} is to integrate these data-driven functions into optimization.

The actual room temperature is difficult to accurately predict due to external disturbances and measurement deviations.
Therefore, real-time data is used to perform iterative optimization and the HVAC operation schedule is updated to minimize the difference between the planning and the actual room temperature.
Moreover, data-driven approaches enable the framework to adaptively predict the environment and system state by learning the thermal dynamics from the data.
For the practical case, after solving the optimization problem, the room temperature obtained in the first time step is used as the set point of the actual thermostat controller.
In this way, our proposed framework controls the actual HVAC system.

\section{Symbolic Regression Based Room Temperature Prediction} \label{sec:SRM}

In the proposed HVAC control framework, the symbolic regression model (SRM) is adopted for the prediction of room temperature.
Compared to predicting room temperature using a thermal equivalent circuit model in \cite{zhao, Ferracuti2017}, the symbolic regression method does not require detailed building information and significantly simplified the parameter identification process.
\begin{figure}[!t]
    \centering
    \includegraphics[width=0.9\linewidth]{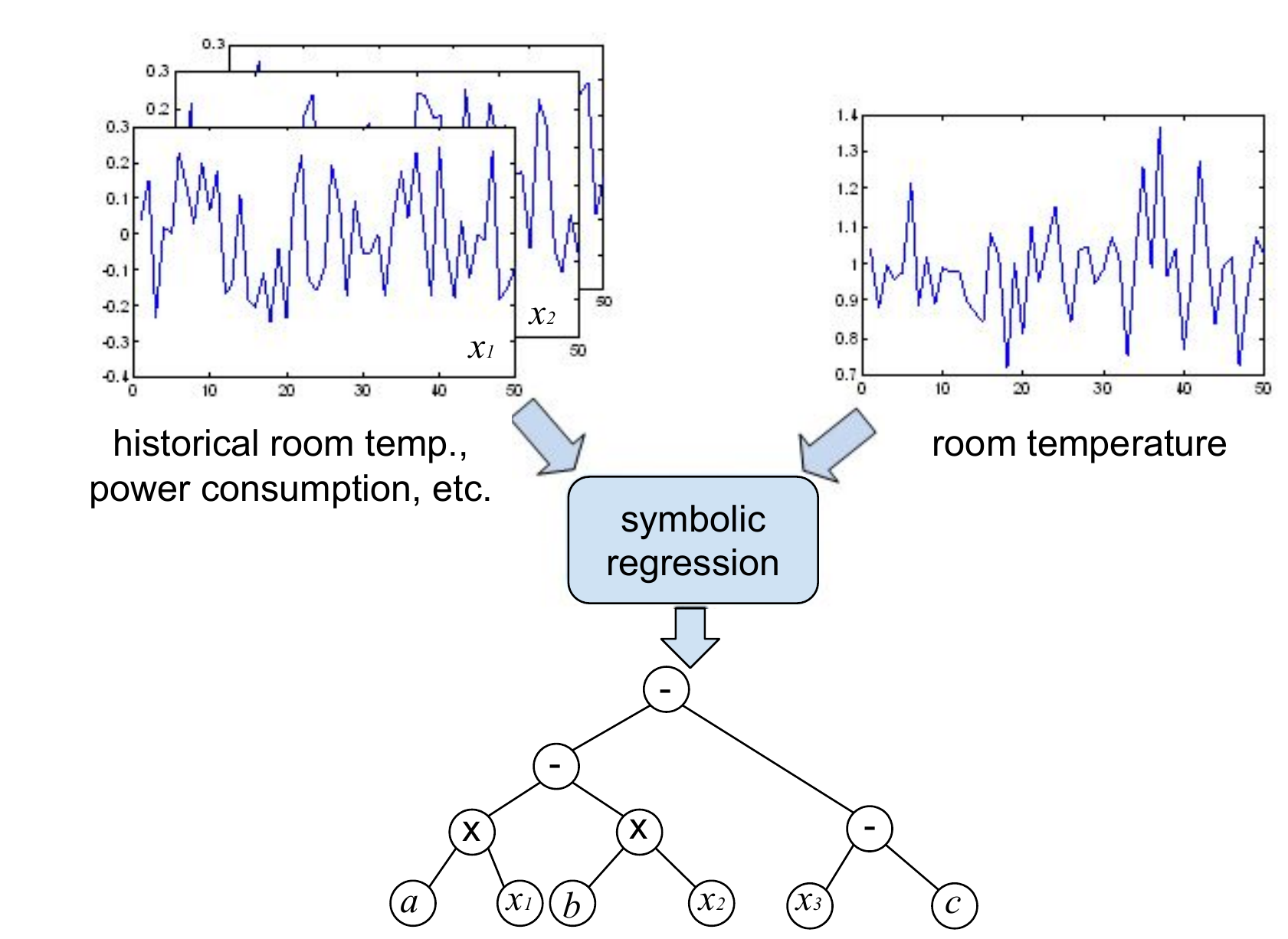}
    \caption{Symbolic regression.}
    \label{fig:sr}
\end{figure}

Symbolic regression is a regression analysis that uses genetic programming to find the most appropriate function that is accurate and simple for a given set of data, as shown in Fig.~\ref{fig:sr}.
For the objective variable of symbolic regression is $Y$ and the explanatory variable is $\bm{X}$, symbolic regression finds combinations of variables, coefficients, and operators of the function $f^{SRM}$ such that $Y=f^{SRM}(\bm{X})$.
In this study, we adopt symbolic regression to reveal the relationship between room temperature at a certain time and historical observation data \cite{Leprince2021}.
The objective variable $Y$ is the room temperature $T^{in}_{t+1}$ at time $t$+1, and the explanatory variables $\bm{X}_t$ are the room temperature $T^{in}_{t}$, ambient temperature $T^{out}_{t}$, power consumption $D_{t}$, HVAC capacity $Q_{t}$, etc. before time $t$, as listed in Table~\ref{tab:explanatory variable}.
Where the historical data contains not only one-time step behind, but also up to $a\cdot n$ steps before with a resolution of $a$ steps.

\begin{table*}[!t] 
\begin{center}
\caption{Example of explanatory variable $\bm{X}_t$}
	\label{tab:explanatory variable}
	\resizebox{\linewidth}{!}{
	\begin{tabular}{c|c|c|c|c|c|c|c|c} \hline
         &\begin{tabular}{c}time t \\(current time)\end{tabular}&\begin{tabular}{c}time $t-1$\end{tabular}&\begin{tabular}{c}time $t-a$\end{tabular}&&\begin{tabular}{c}time $t - a\cdot k$\end{tabular}& &\begin{tabular}{c}time $t-a(n-1)$\end{tabular}&\begin{tabular}{c}time $t-a\cdot n$\end{tabular}\\ \hline  \hline

         %元の
		 %&\begin{tabular}{c}current time\\ (time step $t$) \end{tabular}&\begin{tabular}{c}$1$ step\\before\end{tabular}&\begin{tabular}{c}$a$ step\\before\end{tabular}&&\begin{tabular}{c}$a\cdot k$ step\\before\end{tabular}& &\begin{tabular}{c}$a(n-1)$ step\\before\end{tabular}&\begin{tabular}{c}$a\cdot n$ step\\before\end{tabular}\\ \hline  \hline
   
          \begin{tabular}{c} Room\\temperature\end{tabular} &$T_{t}^{in}$&$T_{t-1}^{in}$&$T_{t-a}^{in}$&$\cdots$&$T_{t-a\cdot k}^{in}$&$\cdots$&$T_{t-a(n-1)}^{in}$&$T_{t-a\cdot n}^{in}$\\ \cline{1-9}
        \begin{tabular}{c} Ambient\\temperature\end{tabular}&$T_{t}^{out}$&$T_{t-1}^{out}$&$T_{t-a}^{out}$&$\cdots$&$T_{t-a\cdot k}^{out}$&$\cdots$&$T_{t-a(n-1)}^{out}$&$T_{t-a\cdot n}^{out}$\\ \cline{1-9}
        %\begin{tabular}{c} AC\\power\end{tabular}&$Q_{t}$&$Q_{t-1}$&$Q_{t-a}$&$\cdots$&$Q_{t-a\cdot k}$&$\cdots$&$Q_{t-a(n-1)}$&$Q_{t-a\cdot n}$\\ \cline{1-9}
        \begin{tabular}{c} Power\\consumption\end{tabular}&$D_{t}$&$D_{t-1}$&$D_{t-a}$&$\cdots$&$D_{t-a\cdot k}$&$\cdots$&$D_{t-a(n-1)}$&$D_{t-a\cdot n}$\\ \cline{1-9}
        $\vdots$&\multicolumn{8}{|c}{\vdots} \\ \hline
	\end{tabular}}
\end{center}
\end{table*}

\section{HVAC model} \label{sec:HVAC}

To estimate the power consumption of the HVAC system under different operating conditions, in this study, we introduced a piecewise linear HVAC model denoted by $f^{HVAC}$.
We assume that the only heat that HVAC can control is sensible heat, and the change in latent heat is considered as a side effect of HVAC operation.
That is, the proposed HVAC model only considers sensible heat capacity such as HVAC heating / cooling capacity and power consumption.

\begin{figure}[!t]
    \centering
    \includegraphics[width=0.9\linewidth]{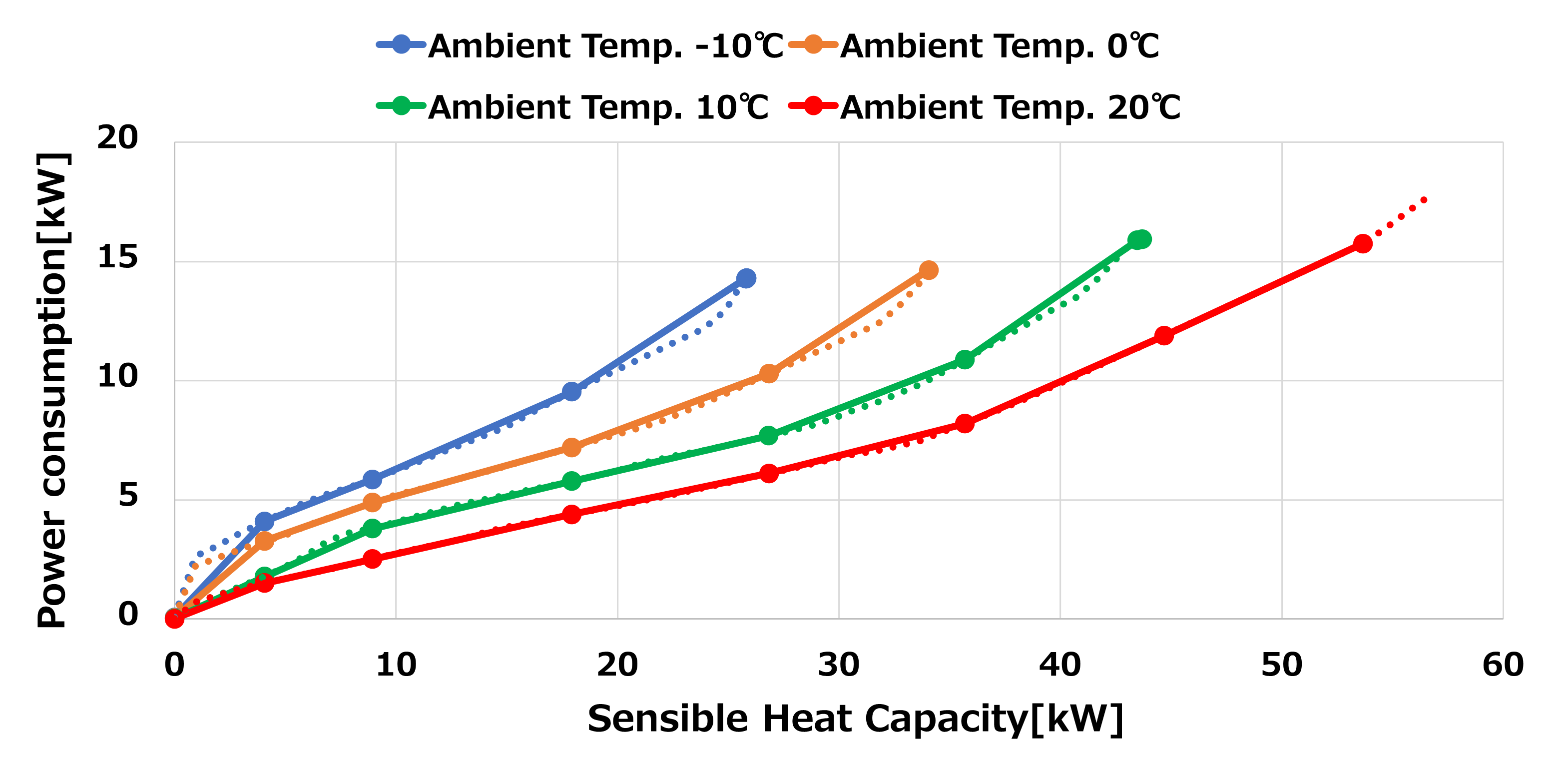}
    \caption{Piecewise linear functions for HVAC model.}
    \label{fig:pwf}
\end{figure}

In this model, the power consumption and heating / cooling capacity of HVAC are expressed by a group of piecewise linear functions.
% created by simulating the operation of the AC and sampling the change in power consumption when the sensible heat load changes.
Since the relationship between HVAC capacity and power consumption is affected by ambient temperature, piecewise linear functions are separated by different ambient temperature levels.
Therefore, the function of HVAC thermal capacity is represented by $f^{HVAC}(T^{out}_{t},T^{in}_{t})$.
% simulations are performed by changing the ambient temperature to create a piecewise linear function for each ambient temperature.
Fig.~\ref{fig:pwf} shows an example of piecewise-linear functions for sensible heat capacity and power consumption.
It assumes a heating operation with one DAIKIN RXYP335FA outdoor unit and four DAIKIN FXYFP80MM indoor units.
Where the blue, orange, green, and yellow dotted lines represent the relationship between sensible heat capacity and power consumption at ambient temperature of -10$^{\circ}$C, 0$^{\circ}$C, 10$^{\circ}$C, and 20$^{\circ}$C, respectively\footnote{Room temperature: 20$^{\circ}$C, Indoor/outdoor humidity: 50\%, indoor latent heat load: 0kW.}.
The solid lines represent the piecewise linear result at different ambient temperature levels.
% The dotted lines represent the sensible heat capacity and power consumption when the indoor temperature is set at 20°C, the outdoor humidity at 50\%, the indoor humidity at 50\%, the latent heat load at 0 kW, and the sensible heat load is varied from 0 kW to 48 kW in 1 kW increments.
% The Piecewise linear function actually used as the HVAC model, with the data points reduced from this fine simulation result, is represented by the solid line.
% Since the maximum sensible heat capacity at each ambient temperature is 26 kW, 34 kW, 44 kW, and 56 kW, the nodes are selected here in 9 kW increments as their approximate common divisors.

Furthermore, since HVAC does not operate when its heating/cooling capacity is below a certain level, constraints are set that HVAC is turned off at HVAC capacities below a threshold.
In the proposed HVAC model, the threshold for turn-off HVAC is set when the sensible heating of the HVAC is below 4.125 kW, assuming that the HVAC starts to operate when the HVAC load is greater than 0.11.
% Here, a node is added to clearly indicate the power consumption at the air-conditioning capacity threshold.
% In this example, a point with a sensible heat capacity of 4.125 kW is selected as a node, assuming that the AC is not operated when the AC load ratio is less than 0.11.
% As described above, simulation of AC is performed while roughly varying the sensible heat load, and the results are used to create a piecewise linear function.
% Using the piecewise linear function created in this way, the relationship between air-conditioning capacity and power consumption is modeled.

\section{On-site experiment} \label{sec:exp}
We implemented the proposed framework in a laboratory at Osaka University to conduct an on-site experiment to evaluate the proposed framework.
The target room is located in the southeast corner of the fourth floor of a six-story building with an area of 179 $m^2$ and a ceiling height of 2.9 $m$.
The target room contains two parts, a work space for 164 $m^2$ and a thermally isolated server room for 15 $m^2$, and the comfort of the work space is considered in the on-site experiment.

\subsection{Experiment setup}
Six HITACHI RCI-GP90K3 indoor units are installed in the target room, which are connected to a single HITACHI RAS-AP560SSR outdoor unit.
Five of the six indoor units are located in the working space and one in the server room; the indoor unit in the server room is not considered in this experiment.

In this implementation, the temperature of the room takes the average value of the intake temperatures from the HVAC indoor units.
Ambient temperature prediction data are obtained from weather forecast data around the Suita campus of Osaka University using OpenWeatherMap.
The proposed framework is performed on the edge computing device\footnote{OpenBlocks IoT VX2, Intel Atom E3805 CPU (dual-core, 1.33 GHz) with 2 GB memory.}, the HVAC setpoints, and the start / stop status determined by the optimization results.

The control and prediction horizons for the optimization calculation are 24 hours each, and the optimization calculation is performed every 15 minutes to update the control input.
Furthermore, the peak power demand and the ramp rate of the HVAC system are strongly suppressed from 5:00 to 10:00.
% The optimization calculation to determine the AC schedule was comfort-oriented. ($\omega = 0.1$).
% In addition, the peaks were more strongly suppressed between 5:00 and 10:00.
% CPLEX v20.1 was used as the MIP solver.

Room temperature and HVAC energy/power consumption were experimentally verified from 0:00 to 24:00 on September 24th and 25th, 2022.
The goal is to maintain the room temperature at 20$^{\circ}$C from 7:00 to 18:00.
For comparison, a reference experiment was conducted on October 1st and 2nd, 2022, from 0:00 to 24:00.
The HVAC system is controlled by a rule-based strategy in which the HVAC setpoint was 20$^{\circ}$C and the operating hours are from 7:00 to 18:00.
% AC was started at 7:00 with a
% set temperature of 20°C and stopped at 18:00. Henceforth, this
% is referred to as manual operation

% For the SRM described in Section \ref{sec:SRM}, the Python library PySR \cite{pysr} was used for the symbolic regression.
Symbolic regression is training on a high-performance server\footnote{PRIMERGY RX4770 M3, Intel(R) Xeon(R) CPU E7-8890 v4 (24 core, 2.20 GHz) with 16 GB memory.} with PySR library~\cite{pysr}.
To simplify the optimization process, the SRM uses only operators with addition, subtraction, and multiplication, so a linear equation is always obtained.
% The operators used in the prediction equation are addition, subtraction, and multiplication, so that a linear equation is always obtained.
The decision variables described in Section~\ref{sec:SRM} are room temperature, HVAC power consumption, and ambient temperature prediction data with $a=12$ (3 hours) and $n=8$, that is, data from 15 minutes, 3, 6... 24 hours before.
Furthermore, to avoid falling into local solutions, symbolic regressions were performed five times under the same conditions, and the best equation was selected from the equations obtained.

% For comparison, an experiment was conducted on Oct. 1 and 2, 2022, from 0:00 to 24:00, with the AC running as usual in the target room.
% The AC was started at 7:00 with a set temperature of 20°C and stopped at 18:00.
% Henceforth, this is referred to as manual operation.

\subsection{Experiment result}
The temperature prediction for the target room on September 24th and 25th was extracted based on historical data as Eq.~\ref{eq:24} and Eq.~\ref{eq:25}, respectively.
\begin{equation} \label{eq:24}
     T_{t+1}^{in} = 0.754 \cdot T_t^{in} - 0.162 \cdot T_{t-60}^{in} - D_{t} + 10.445
\end{equation}
\begin{equation} \label{eq:25}
     T_{t+1}^{in} = 0.792 \cdot T_t^{in} - 0.718 \cdot D_t - 0.103 \cdot T_{t-24}^{out} + 6.290
\end{equation}
where $T^{in}$, $T^{out}$, and $D_{t}$ are the room temperature, the ambient temperature, and the power consumption of the HVAC system, respectively.

\begin{figure}[!t]
    \centering
    \includegraphics[width=0.9\linewidth]{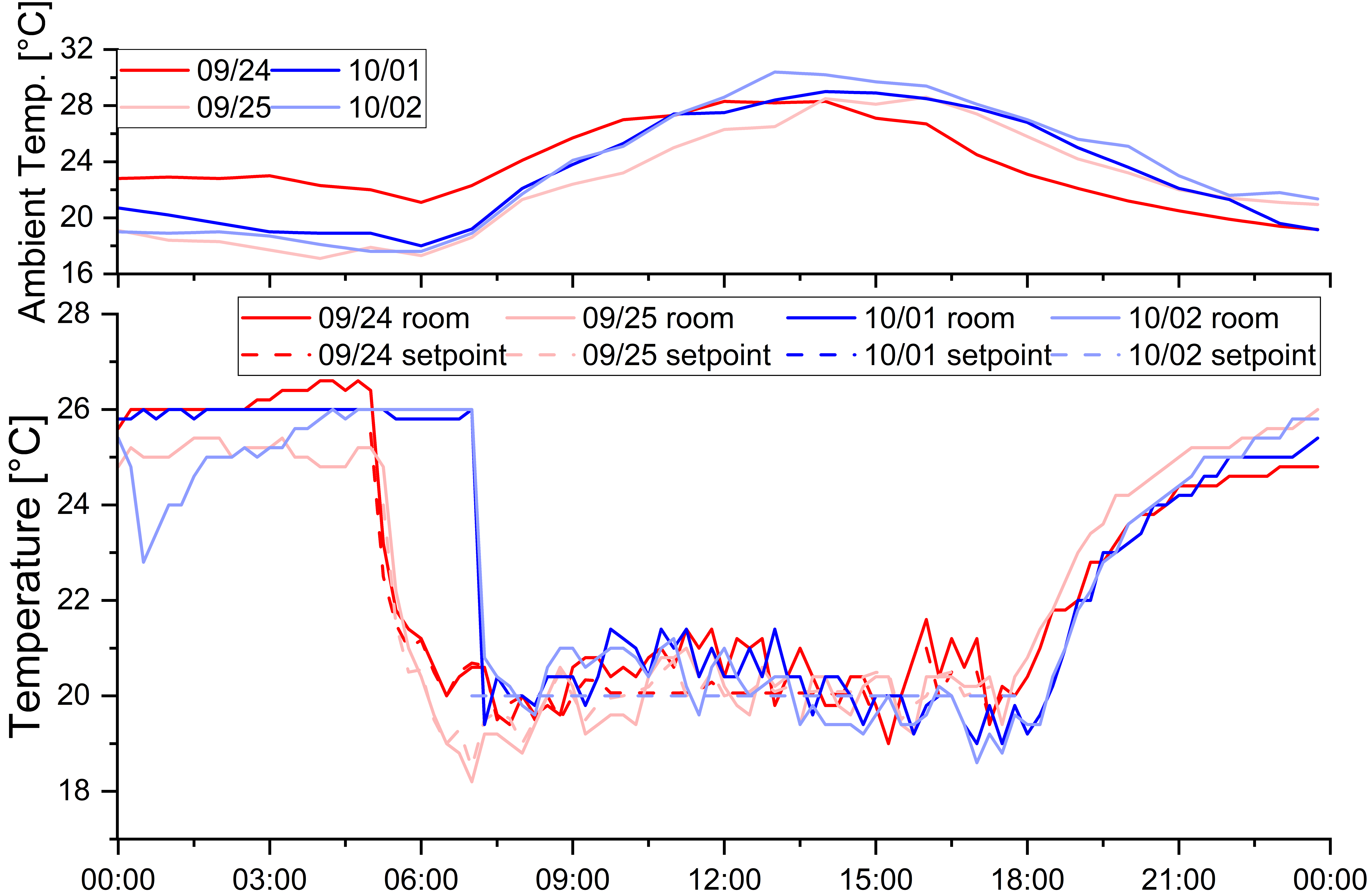}
    \caption{Experimental results: temperature variation.}
    \label{fig:result_temp}
\end{figure}

\begin{figure}[!t]
    \centering
    \includegraphics[width=0.9\linewidth]{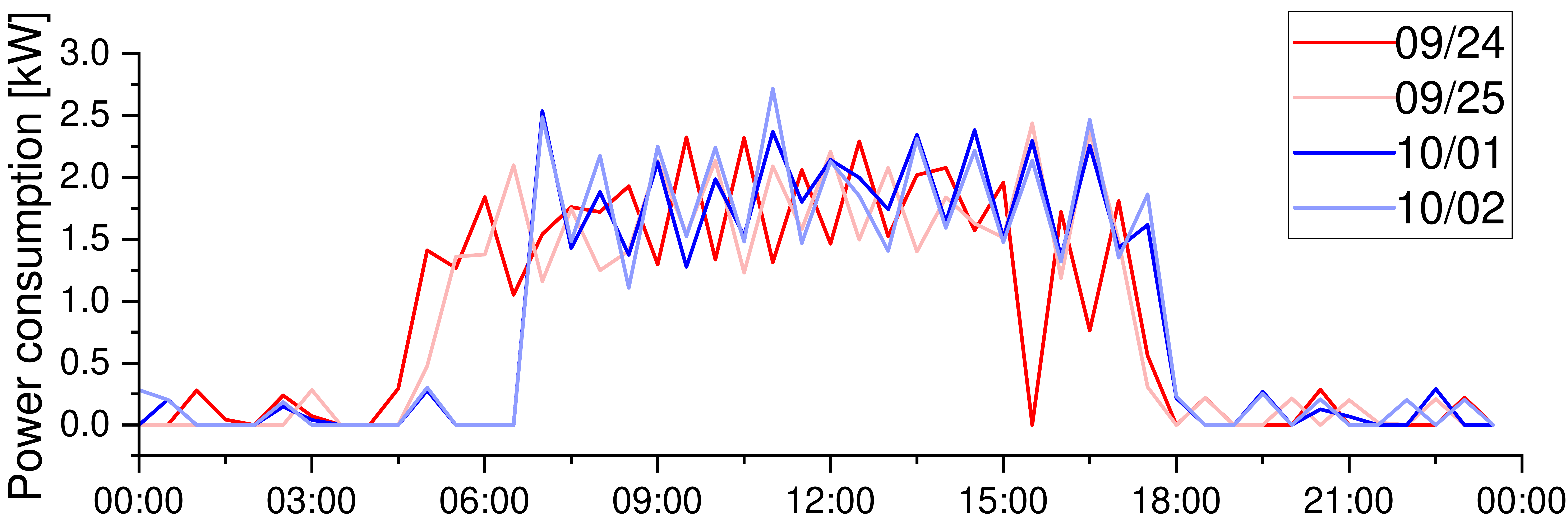}
    \caption{Experimental results: power consumption.}
    \label{fig:result_demand}
\end{figure}

\begin{figure}[!t]
    \centering
    \includegraphics[width=0.9\linewidth]{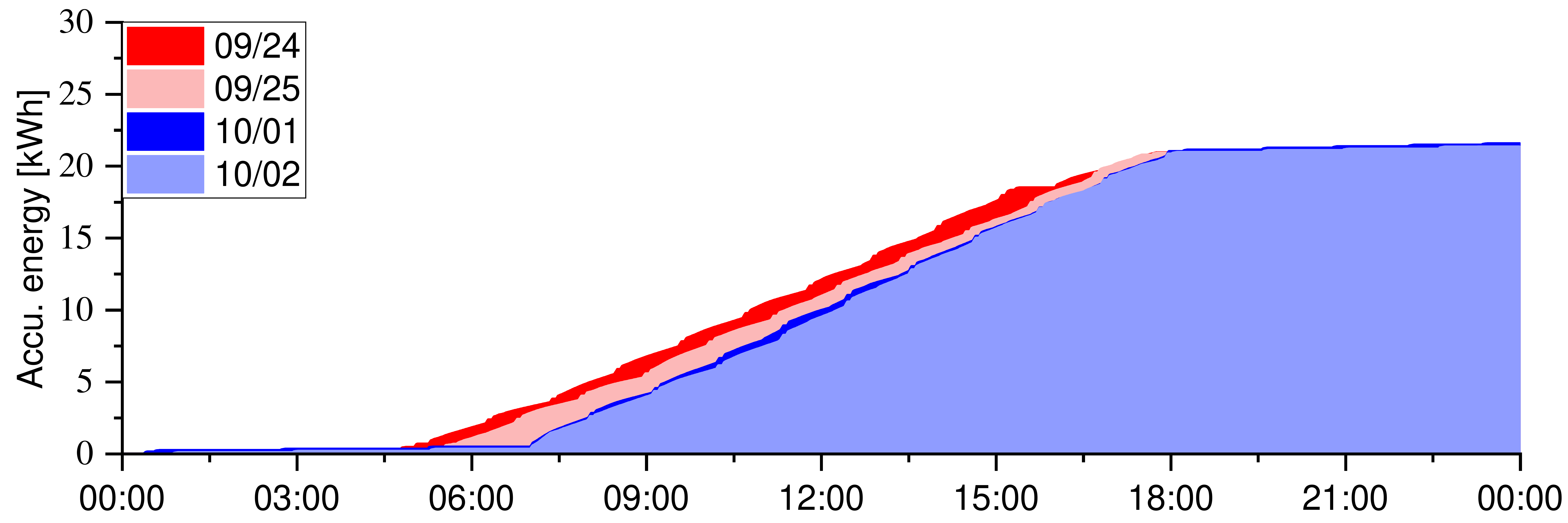}
    \caption{Experimental results: accumulated energy consumption.}
    \label{fig:result_acc}
\end{figure}
Fig.~\ref{fig:result_temp} shows the variation in ambient and room temperature in the target room during the experiment.
% The thick and thin red lines show the results for September 24 and 25, respectively, and these are the results of the proposed method.
% The thick and thin blue lines indicate Oct. 1 and Oct. 2, respectively, and these are the results obtained using the manual operation.
As can be seen in Fig.~\ref{fig:result_temp}, the room temperature with the rule-based control strategy started to decrease at 7:00 (dark and light blue curve), while with the proposed framework being performed, the room temperature started to decrease gently from 5:00 to precool the target room (dark and light red curve).
% In the graph of temperatures in Fig.\ref{fig:result_temp}, the solid blue line shows that the room temperature in the case of manual operation started to decrease at 7:00 a.m., while the red line shows that the room temperature started to decrease slowly at 5:00 a.m. when the proposed method was used, which confirms that precooling was performed by the proposed method. This confirms that precooling is performed by the proposed method.

Fig.~\ref{fig:result_demand} shows the power consumption at the target during the experiment.
As can be seen in Fig.~\ref{fig:result_demand}, the power consumption with the rule-based control strategy started to increase at 7:00 (dark and light blue curve), while with the proposed framework, the power consumption started to increase gently from 5:00 to precool the target room (dark and light red curve).

Fig.~\ref{fig:result_demand} and ~\ref{fig:result_acc} show the power and accumulated energy consumption in the target during the experiment.
When the proposed framework was not used, the HVAC system worked in full power mode from the beginning of its operation, rapidly increasing its workload.
%While the total energy consumption in the manual operation shown in blue and light blue started to increase from 7:00, the total energy consumption in the proposed method shown in red and light red started to increase from around 5:00.
%In addition, the increase in the totalization value in the proposed method ends earlier than 18:00, which means that the operation is not performed during the time period when it is not necessary to achieve the target temperature.
As can be seen in Fig.\ref{fig:result_acc}, the accumulated energy consumption with the rule-based control strategy started to increase at 7:00 (dark and light blue area), while with the proposed framework, the power consumption started to increase gently from 5:00.
Furthermore, the increase in accumulated energy consumption with the proposed framework ends earlier than 18:00, since the temperature setpoint has been reached.
% This means that the AC had stopped during the time period when it is not necessary to achieve the target temperature.
% This means that AC stopped during times when there is little need to achieve the target temperature.

\begin{table}[!tbp] 
\begin{center}
\caption{On-site experiment result}
\vspace{-15pt}
	\label{tab:result}
	\resizebox{\linewidth}{!}{
	\begin{tabular}{c|c|c|c} \hline
        date&method&\begin{tabular}{c}Peak\\power\\ \lbrack kW\rbrack \end{tabular}&\begin{tabular}{c}Energy\\consumption\\ \lbrack kWh\rbrack \end{tabular}\\ \hline \hline 
        09/24& Proposed & 2.32 & 21.3 \\ \hline
        09/25& Proposed & 2.13 & 21.3 \\ \hline \hline
        10/01& Rule-based & 2.54 & 21.3  \\ \hline
        10/02& Rule-based & 2.49 & 21.6 \\ \hline 
	\end{tabular}}
\end{center}
\end{table}

Table \ref{tab:result} summarized the 30-minute averaged peak power demand (the maximum power consumption between 5:00 and 10:00) and the total energy consumption during the experiment.
% shows the operation method, 30-minute average peak power demand (the maximum power consumption between 5:00 and 10:00), total power consumption and maximum temperature, and optimum temperature for each experimental day.
Compared to the rule-based control strategy, the proposed method reduces the peak power demand by a maximum of 16.1 \%, and the total energy consumption maintains the same level and when pre-cooling is performed.
% is almost the same despite the precooling by the proposed method.
These results indicate that the proposed method can reduce the peak power demand with pre-cooling performed while maintaining the same level of total energy consumption.

\section{Conclusion} \label{sec:conclusion}
In this paper, we have presented the design and implementation of an online HVAC optimization framework that incorporates a room temperature prediction model using symbolic regression to achieve thermal comfort and reduce peak power demand.
In addition, a piecewise linear function-based HVAC model is proposed to capture the relationship between HVAC capacity and power consumption.
% was created by simulating AC units and sampling the results, and a model was proposed to represent the relationship between AC capacity and power consumption through this function.
%The evaluation was performed with measured data in the laboratory and showed high accuracy with an MSE of 0.57 for measured and predicted room temperatures.
The results of the on-site experiment demonstrated that the proposed framework can reduce the peak power demand by up to 16.1 \% with almost no change in total energy consumption and maintain reasonable thermal comfort level.
% laboratory demonstrations show that the proposed framework can reduce peak power demand by up to 16.1 \% with almost no change in total power consumption.
In the future, we intend to study the parameters of symbolic regression and the amount of input data in more detail to achieve more accurate predictions and to enable symbolic regression that can be performed in parallel with optimization calculations on edge servers.
We also aim to introduce room occupancy estimation and prediction into the framework for occupancy-driven HVAC control.

\bibliographystyle{IEEEtran}
\bibliography{name.bib}

\end{document}